\documentstyle[12pt]{article}

\def\ltsim{\lower3pt\hbox{$\, \buildrel < \over \sim \, $}}
\def\gtsim{\lower3pt\hbox{$\, \buildrel > \over \sim \, $}}
\def\be{\begin{equation}}
\def\ee{\end{equation}}
\def\ba{\begin{eqnarray}}
\def\ea{\end{eqnarray}}
\def\bea{\begin{eqnarray}}   
\def\eea{\end{eqnarray}}

\def\ga{\mathrel{\raise.3ex\hbox{$>$\kern-.75em\lower1ex\hbox{$\sim$}}}}
\def\la{\mathrel{\raise.3ex\hbox{$<$\kern-.75em\lower1ex\hbox{$\sim$}}}}

\begin{document}

\baselineskip=16pt
\begin{titlepage}
\rightline{OUTP-02/44P}
\rightline{HIP-2002-65/TH}
\rightline{hep-th/0212137}
\rightline{ December 2002}
\begin{center}

\vspace{1cm}

\large {\bf  GHOST-MATTER MIXING AND FEIGENBAUM UNIVERSALITY IN STRING 
THEORY}

\vspace{1 cm}
\normalsize

\centerline{
Ian I. Kogan\footnote{i.kogan@physics.ox.ac.uk}$^{\,a,b}$  and
 Dimitri Polyakov\footnote{ polyakov@pcu.helsinki.fi}$^{\,c}$
}
\smallskip
\medskip
$^a$
{\em Theoretical Physics, Department of Physics, Oxford University,
 1 Keble Road, Oxford, OX1 3NP,  UK}\\
$^b$ {\em ITEP, Moscow, Russia} \\
$^c$ {\em Department of Physical Sciences, University of Helsinki
and Helsinki Institute of Physics, PL 64, FIN-00014 Helsinki, Finland}
\smallskip

\vskip0.2in 

{\it Dedicated
 to the 80-th birthday of Karen Avetovich Ter-Martirosyan.}
\end{center}
\vskip0.2in 

\centerline{\large\bf Abstract}

Brane-like vertex operators, defining  backgrounds with the
ghost-matter mixing in  Neveu-Schwarz-Ramond (NSR)  superstring theory,
play an important role in a world-sheet 
formulation of D-branes and M theory, being creation operators for 
extended objects in the second quantized formalism.
In this paper we show that  dilaton's beta function in ghost-matter mixing 
backgrounds becomes stochastic. The renormalization group (RG) equations
in ghost-matter mixing backgrounds lead to non-Markovian Fokker-Planck
equations which solutions describe superstrings in curved space-times
with brane-like metrics.We show that Feigenbaum universality constant
$\delta=4,669...$ describing transitions from order to chaos in a huge
variety of dynamical systems, appears analytically in these RG equations.
We find that the appearance of this constant is related to the scaling 
of relative space-time curvatures at fixed points of the RG flow.
In this picture the fixed points correspond to the period doubling
of Feigenbaum iterational schemes. 

\vspace*{2mm}

\end{titlepage}

\section{Introduction}

Superstring theory is our current hope to put gravity in a 
 Prokrust's bed of quantum mechanics. In spite of all the 
spectacular progress in the last quarter of the  century \cite{books}
 the  full structure and underlying symmetries of the theory  
have yet to be unveiled.  One of the most striking features
 of String theory is a deep  relation between 
 renormalization group (RG) flows on a world sheet and an 
evolution in a target space. 
 Critical points of these RG flows, described by  
2d conformal field theories (CFT),
 determine equations of motion in a target space. 
The structure of these equations is determined by the
world sheet correlation functions of the appropriate
vertex operators in  respective CFT \cite{rg}
Thus in the standard string perturbation theory 
the beta beta-function equations describe 
the behaviour of small fluctuations around flat
 backgrounds. The conformal field theory description
of strings in curved backgrounds, such as of strings 
in the presence of branes, as well as the 
underlying CFT of strongly coupled strings is much harder
a problem to tackle, in particular because the adequate  knowledge
of quantum degrees of freedom of M-theory and non-perturbative
strings is still lacking.
Some time ago, we have proposed the formalism 
\cite{Polyakov:2000xc}-\cite{Kogan:2002tn}
that describes the non-perturbative dynamics of solitons in
string and M-theory in terms of a special class of 
vertex operators, called brane-like states.
The crucial distinction of these vertex operators from usual one
(such as a photon or a graviton) is that they exist at nonzero
ghost pictures only. The simplest example of these vertices in the
closed string case is given by:
\bea
V_5^{(-3)}(q)=\int{d^2z}e^{-3\phi-{\bar\phi}}\psi_{t_1}...\psi_{t_5}
{\bar\psi}_{t_6}e^{iq_a{X_a}}(z,{\bar{z}})
\nonumber \\
V_5^{(-2)}(q)= \int{d^2z}c{\partial\chi}
e^{\chi-3\phi-{\bar\phi}}\psi_{t_1}...\psi_{t_5}
{\bar\psi}_{t_6}e^{iq_a{X_a}}(z,{\bar{z}})
\nonumber \\
V_5^{(+1)}(q)= \int{d^2z}e^{\phi-{\bar\phi}}\psi_{t_1}...\psi_{t_5}
{\bar\psi}_{t_6}e^{iq_a{X_a}}(z,{\bar{z}})+b-c ~~ghosts
\nonumber \\
a=0,...,3;~~~~~~~~t_i=4,...9; ~~~~~~~~~~
\eea
It is important that the BRST nontriviality and invariance conditions
on this vertex  confine its propagation to the 4-dimensional subspace
(labeled by the $a$ index), transverse to its polarization defined by 
6 $t_i$ indices.
This six-form vertex exists only at pictures below $-2$ or above $+1$
while there is no version of this operator at pictures $0$  and $-1$.

 This means that the discrete picture-changing gauge
symmetry
is broken for such operators and their superconformal ghost dependence
cannot be removed
by any picture-changing transformation. We shall refer to this property of
the brane-like 
vertices as the ghost-matter mixing.
The crucial property of these special vertex operators
is that  they do not correspond to any perturbative string excitation but
describe the nonperturbative dynamics of extended solitonic objects, such
as D-branes.
It appears that the non-perturbative character of these vertices is
closely
related to the their ghost-matter mixing properties on the world sheet,
with the latter
encoding the crucial information about the brane dynamics.
 In  \cite{Kogan:2002tn} we have shown that the low-energy effective
action of the sigma-model with the brane-like states is given by the
Dirac-Born-Infeld (DBI) action for D-branes. From the world sheet point of 
view this means 
that 
the insertion of vertices with the ghost-matter mixing makes the
 deform  CFT describing
 strings in flat space-time  and it flows to  a new fixed point,
corresponding 
to the CFT of strings in a curved background  induced by D-branes.
In this paper we shall further investigate  RG flows in the 
ghost-matter mixing backgrounds. It appears that properties of these RG
flows are 
stunningly different from the usual ones.   We  found that 
ghost-matter mixing  adds to RG flow operator-valued stochastic terms.
Even more intriguing is the  emergence of universal constant in the RG
equations
 which with accuracy less than $0.5 \% $ is 
 nothing but the logarithm of the famous Feigenbaum constant $\delta =
4.669$ \cite{Feigenbaum:ys}. 
 This coincidence is not accidental but  reflects  remarkable and new
relations  between superstrings, chaos, gravity and stochastic processes
 which is the subject of this Letter.

\section{Dilaton beta-function in ghost-matter mixing backgrounds}

The crucial property of  world sheet conformal beta-functions
(e.g. of a dilaton) 
in ghost-matter mixing backgrounds is the presence of stochastic terms
in the RG equations. In usual perturbative backgrounds
(e.g. of a graviton or an axion) such terms are  absent
and the beta-function equations are deterministic.
Things, however, become different in backgrounds with the
ghost-matter mixing.
One specific property of the brane-like states, distinguishing
them from usual perturbative vertex operators, is that their OPE 
algebra is picture-dependent. Below we will show that this
picture dependence leads to non-deterministic stochastic terms
in the dilaton's beta-function. The stochasticity of the renormalization
group, in turn, will be shown to be closely related to the nonperturbative
nature of the brane-like states and their relevance to the 
non-perturbative description of branes and strings in curved backgrounds.
Let us start with the Neveu-Schwarz-Ramond (NSR) sigma model in $D=10$ 
perturbed by the 
dilaton and
the ghost-matter mixing vertex (1).
The generating functional for this model is 
given by:
\bea
Z(\varphi,\lambda)=\int{DX}D\psi{D}{\lbrack{ghosts}\rbrack}:f(\Gamma):   
:f(\bar\Gamma):
\nonumber \\
exp{\lbrace}{-S_{NSR}}+\int{d^4q}\lambda(q)\int{d^2z}V_5^{(-3)}(q,z,\bar{z})
+\int{d^{10}p}\varphi(p)\int{d^2w}
V_\varphi^{(-2)}(p,w,\bar{w}){\rbrace} 
\eea
Here
$V_5^{(-3)}$ is the ghost-matter mixing vertex (1), creating
the D3-brane background, $V_\varphi^{(-2)}$ is the dilaton vertex operator
taken at the  ghost picture $-2$.
$$:f(\Gamma):=:{1\over{1-\Gamma}}:=1+:\Gamma:+:\Gamma^2:+...$$
is the measure function of  picture-changing operator,
$:\Gamma:=:e^\phi{G}:$ with $G=G_m+G_{gh}$ being the full
matter $+$ ghost world sheet supercurrent. This measure function is
necessary to insure the correct ghost number balance in correlation 
function; $:f(\bar\Gamma):$ is defined similarly.
In the perturbative case the $:f(\Gamma){{f}}({\bar{\Gamma}}):$
insertion simply insures that all the operator product expansions (OPE) 
of the vertex operators
are equivalent at all the picture levels and all the structure
correlation functions are picture independent.
The dilaton vertex operator can be taken at any negative picture.
The negative value of the dilaton picture, along with the $:f(\Gamma):$
in the measure, insures the correct overall ghost number in  correlation 
functions involving the dilaton (which must be equal to -2 on the sphere)
this vertex operator is picture independent (meaning the picture
independence its
matrix elements with other states) and the results of the dilaton's
beta-function are independent on the $V_\varphi$'s picture as well.
It is convenient to take $V_\varphi$ at picture $-2$ (both left and 
right)
as in this case the dilaton vertex operator is given by
$$V_\varphi(p)=\int{d^2z}e^{-2\phi-2\bar\phi}\partial{X^m}\partial{X_n}
(\eta_{mn}-k_m{\bar{k}}_n-k_n{\bar{k}}_m)$$.
In the absence of ghost-matter mixing the functions
$:f(\Gamma):$ and $:f(\bar\Gamma):$ simply insure that
the beta-function equations for any closed string
space-time fields are identical for any 
picture representation
of corresponding closed string vertex operators.
In order to compute the dilaton's beta-function in the presence
of the $V_5$-operator, one has to expand this generating functional
in $\varphi$ and $\lambda$. In the absence of ghost-matter mixing,
the dilaton's (or any other closed string field's) 
beta-function in any background is simply given by the appropriate
three-point correlators, or the structure constants.
In the ghost-matter mixing case, however, the situation is different
since the structure constants are picture-dependent and therefore
expressing the beta-function in terms of three-point correlation 
functions is not well-defined.
Therefore to determine  the beta-function one has to 
point out the UV logarithmic divergences due to the OPE singularities
(generally speaking, picture dependent)
and to sum over all the pictures.

Expanding the generating functional (2) 
up to the third order  of $\lambda$ and the second order of $\varphi$
we obtain:
\bea
Z(\varphi,\lambda)=\int{DX}D\psi{D{\lbrack{ghosts}\rbrack}}:f(\Gamma):
:f(\bar\Gamma): 
e^{{-S_{NSR}}}
\nonumber \\
\lbrace
1+{\int}d^{10}p\varphi(p)\int{d^2z}V_\varphi^{(-2)}(z,\bar{z};p)
+\int{d^4q}\lambda(q)\int{d^2w}V_5^{(-3)}(w,\bar{w};q)
\nonumber \\
+{\int}d^{10}p{\int}d^{4}q\lambda(q)\varphi(p)\int{d^2z}\int{d^2w}
V_\varphi^{(-2)}(p;z,\bar{z})V_5^{(-3)}(w,\bar{w};q)
\nonumber \\
+{1\over2}\int{d^4q_1}\lambda(q_1)\int{d^4q_2}\lambda(q_2)
\int{d^2w_1}\int{d^2w_2}
V_5^{(-3)}(w_1,\bar{w_1};q_1)V_5^{(-3)}(w_2,\bar{w_2};q_2)
\nonumber \\
+{1\over2}\int{d^{10}p_1}\varphi(p_1)\int{d^{10}p_2}\varphi(p_2)
\int{d^2z_1}\int{d^2z_2}
V_\varphi^{(-2)}(z_1,\bar{z_1};p_1)V_\varphi^{(-2)}(z_2,\bar{z_2};p_2)
\nonumber \\
+{1\over6}\int{d^4q_1}\int{d^4q_2}\int{d^4q_3}\lambda(q_1)\lambda(q_2)
\lambda(q_3)
\nonumber \\
\times\int{d^2w_1}\int{d^2w_2}\int{d^2w_3}
V_5^{(-3)}(w_1,\bar{w_1};q_1)V_5^{(-3)}(w_2,\bar{w_2};q_2)
V_5^{(-3)}(w_3,\bar{w_3};q_3)+...\rbrace
\eea
where we dropped the higher order terms as well as those irrelevant
to our discussion (such as those of the order of $\lambda^2\varphi$
or $\varphi^2\lambda$).
To determine the UV divergences in the partition function (2),(3),
relevant to the dilaton's beta-function, one has to point out
the relevant singular terms in the OPE algebra of 
 the dilaton and $V_5$. 
In the on-shell limit, the relevant terms in the operator algebra are 
given by:
\bea
V_5^{(-3)}(w_1,\bar{w_1};q_1)V_5^{(-3)}(w_2,\bar{w_2};q_2)
\sim{{C_{{\lbrack}-3|-3\rbrack}(q_1,q_2)
V_\varphi^{(-6)}(q_1+q_2)}\over{|w_1-w_2|^2}}+...
\nonumber \\
V_5^{(-3)}(w_1,\bar{w_1};q_1)V_5^{(+1)}(w_2,\bar{w_2};q_2) 
\sim{{C_{{\lbrack}-3|1\rbrack}(q_1,q_2)
V_\varphi^{(-2)}(q_1+q_2)}\over{|w_1-w_2|^2}}+...
\nonumber \\
V_5^{(+1)}(w_1,\bar{w_1};q_1)V_5^{(+1)}(w_2,\bar{w_2};q_2) 
\sim{{C_{{\lbrack}1|1\rbrack}(q_1,q_2)
V_\varphi^{(+2)}(q_1+q_2)}\over{|w_1-w_2|^2}}+...
\eea
where
\bea
{C_{{\lbrack}-3|-3\rbrack}}(q_1,q_2)\sim(q_1 q_2)(1+(q_1+q_2)^2)
\nonumber \\
{C_{{\lbrack}-3|1\rbrack}}(q_1,q_2)\sim(q_1 q_2)
\nonumber \\
{C_{{\lbrack}-3|1\rbrack}}(q_1,q_2)\sim(q_1 q_2)(1-(q_1+q_2)^2)  
\eea
Next, one has to point out the picture changing rules for
  the left (holomorphic) part of the $V_5$-operator,
in order to specify how it is acted on by $:f(\Gamma):$
The antiholomorphic part and its interaction with
$:f(\bar\Gamma):$ do not interest us since it is the photon-like
part existing at all pictures and all its picture changings are trivial.
The picture changing transformation rules for the $V_5$ operators (1)
can be written in the form
\bea
:\Gamma:^{n}V_5^{(k)}(p)=\alpha_{{\lbrack}k|n+k\rbrack}V_{5}^{(N+k)}(p)
~~~~~~~~~~~~~~~~~~~~~~~~~~~~~~~~~
\nonumber \\
\alpha_{{\lbrack}i|j\rbrack}=
\alpha_{{\lbrack}m|n\rbrack}=\alpha_{{\lbrack}s|t\rbrack} = 1 ; 
~~~~
\alpha_{{\lbrack}a|j\rbrack} =\alpha_{{\lbrack}a|b\rbrack}=  
\alpha_{{\lbrack}s|a\rbrack}=0, ~~~~~
\alpha_{{\lbrack}i|m\rbrack}=\alpha_{{\lbrack}s|m\rbrack}=1+p^2,
\nonumber \\
s,t=-\infty,....-4; ~~~
i,j=-3,-2; ~~~~
a,b=-1,0; ~~~~
m,n=1,2,.... 
\eea
In the beta-function calculations, when the vertex operators
are taken just slightly off-shell,
the following identities are useful:
\bea
\alpha_{{\lbrack}i|m\rbrack}C_{{\lbrack}m|n\rbrack}
=C_{{\lbrack}i|n\rbrack}; ~~~~~
\alpha_{{\lbrack}i|m\rbrack}C_{{\lbrack}m|j\rbrack}           
=C_{{\lbrack}i|j\rbrack}
\eea

Finally, using the fact that picture changing
operators form the polynomial ring:
\be
:\Gamma:^{m+n}=:\Gamma^m\Gamma^n:+\lbrack{Q_{BRST},...}\rbrack
\ee

 the action of the $:\Gamma^n:$ operator 
on the vertex operators inside the functional integral
can be expressed as:
\bea
<:\Gamma^n:(w)V_1(z_1)....V_N(z_N)>
=\sum_{k_1,...k_{N-1}=0}^{k_1+...+k_{N-1}=n}
N^{-n}{{n!}\over{k_1!...k_{N-1}!(n-k_1-...-k_{N-1})!}}
\nonumber \\
\times
<:\Gamma:^{k_1}V_1(z_1)...:\Gamma:^{k_{N-1}}V_{N-1}(z_{N-1})
:\Gamma:^{n-k_1-...-k_{N}}V_N(z_N)>
\eea
i.e. the correlator does not depend on $w$.
The factor of $N^{-n}$ in (9) insures the  correct normalization
of amplitudes in the picture-independent case.
Using the relations (6)-(9) we are finally in the position
to start evaluating the beta-function.
The first contribution of interest to the beta-function
comes from the $\lambda^2$-term, bilinear in the $V_5$-operator.
At any given picture level $n$ this term leads to the following divergence
in the order of $\lambda^2$:
\bea
{1\over2}\int{d^2w_1}\int{d^2w_2}
<:\Gamma^{n+6}: V_5^{(-3)}(w_1,\bar{w_1})
V_5^{(-3)}(w_2,\bar{w_2})...> =
\nonumber \\
=log\Lambda\times{2^{-n-7}}\sum_{k=0}^{n+6} {{(n+6)!}\over{k!(n+6-k)!}}
\alpha_{{\lbrack}-3|k-3\rbrack}C_{{\lbrack}k-3|n+3-k\rbrack} 
\alpha_{{\lbrack}-3|n+3-k\rbrack}
\int{d^2\xi}< {V_\varphi^{(n)}}(\xi,{\bar\xi})...>
\eea
where we introduced the coordinate change:
$\xi=1/2(w_1+w_2),\eta=1/2(w_1-w_2)$ and $log\Lambda=
{\int_\Lambda}
{{d^2\eta}\over{|\eta|^2}}$ is the log of the world sheet UV cutoff,
appearing as a result of the integration over $\eta$
(note that the vertex operators on the r.h.s. of the
operator product (10) are $\eta$-independent)
For the sake of brevity we suppress the momentum dependence
of fields, vertices and structure constants
 here and below.
This divergence is removed by renormalizing the dilaton field as
\bea
\varphi\rightarrow\varphi-{\sum_{n=0}^\infty}
{2^{-n-7}}\sum_{k=0}^{n+6} {{(n+6)!}\over{k!(n+6-k)!}}
\alpha_{{\lbrack}-3|k-3\rbrack}C_{{\lbrack}k-3|n+3-k\rbrack}
\alpha_{{\lbrack}-3|n+3-k\rbrack} 
\lambda^2log\Lambda.
\eea
In the absence of picture-dependence 
the sum over k would have been
reduced to ${{1\over2}C\lambda^2log\Lambda}$ for each picture,
as it should be in the standard case when ghost-matter mixing is absent
(where C are the structure constants with the picture indices 
suppressed)

As a result of the dilaton's  RG flow, the $\lambda\varphi$
cross-term is   renormalized by $\lambda^3$ logarithmic divergence  as:
\bea
\lambda\varphi\int{d^2w_1}\int{d^2w_2}
< :f(\Gamma):V_5^{(-3)}V_\varphi^{(-2)}.. >
\nonumber \\
\lambda^3{log}\Lambda\sum_{n=0}^{\infty}
2^{-2n-12}\sum_{k,l=0}^{k+l=n+5}{{(n+5)!(n+6)!}\over{k!l!(n+5-l)!(n+6-k)!}}
\alpha_{{\lbrack}-3|n+2-l\rbrack}\alpha_{{\lbrack}-3|k-3\rbrack}
\nonumber \\
\times
\alpha_{{\lbrack}-3|n+3-k\rbrack}C_{{\lbrack}k-3|n+3-k\rbrack}
\int{d^2w_1}\int{d^2w_2} <V_\varphi^{(-6)}(w_1,\bar{w_1})
V_5^{(n+6)}(w_2, \bar{w_2})...>
\eea
where in our derivation we used the invariance
of $V_\varphi$ and transformation properties (6) of $V_5$
under the picture-changing.
In the absence of the ghost-matter mixing when
one always has $\alpha=1$ and picture-independent $C$
it is easy to check that the  $\lambda\varphi$
would have been renormalized to 
\be
-{1\over2}C\lambda^3log\Lambda
\int{d^2w_1}\int{d^2w_2} V(w_1,\bar{w_1})
V(w_2,\bar{w_2})
\ee
at each picture level (again, we suppressed the relevant
indices for the structure constants and vertex operators)
In the case under consideration, however,
 using the identities (7) relating $\alpha$ and $C$,
we can cast the renormalization of the $\lambda\varphi$-term 
under the flow (11) as
\bea
A_{\lambda\varphi}\sim
\lambda\varphi:f(\Gamma):\lambda\varphi\int{d^2w_1}\int{d^2w_2} 
V_5^{(-3)}V_\varphi^{(-2)}\rightarrow
\nonumber \\
-\lambda^3{log}\Lambda{C_{{\lbrack}-3|-3\rbrack}}
\alpha_{{\lbrack}-3|1\rbrack}\int{d^2w_1}
\int{d^2w_2}V_\varphi^{(-6)}(w_1,\bar{w_1})
V_5^{(n+6)}(w_2, \bar{w_2})
\nonumber \\ 
\times 
\sum_{n=0}^{\infty}
\sum_{k,l=0;k{\neq}2,3,n+3,n+4;l{\neq}n+2,n+3}^{k=n+6;l=n+5}
{{(n+5)!(n+6)!}\over{k!l!(n+5-l)!(n+6-k)!}}
\eea
This gives the renormalization 
of the $\lambda\varphi$ cross-term under the RG flow (11)
of the dilaton field in the ghost-matter mixing case.
This contribution to the world sheet renormalization group 
has the order of $\lambda^3$ 
The other contribution of the same order of $\lambda^3$ 
to the dilaton beta-function
comes  from the OPE singularities inside the $\lambda^3$-term
itself, appearing in the expansion (3) of the partition function
Using  the OPE (4) and evaluating the singular
world sheet integrals as in (10), (11) we get
\bea
A_{\lambda^3}=
{1\over6}\lambda^3\int{d^2w_1}\int{d^2w_2}\int{d^2w_3}
<V_5^{(-3)}(w_1,{\bar{w_1}})V_5^{(-3)}(w_2,{\bar{w_2}})
V_5^{(-3)}(w_3,{\bar{w_3}})...>
\nonumber \\
\sim
{1\over2}\lambda^3{log}\Lambda{\sum_{n=0}^{\infty}}
3^{-n-10}\sum_{k,l=0}^{k+l=n+9}
{{(n+9)!}\over{k!l!(n+9-k-l)!}}
\nonumber \\ 
\times\int{d^2w_1}\int{d^2w_2} 
<V_\varphi^{(-6)}(w_1,{\bar{w_1}})V_5^{(n+6)}(w_2,
{\bar{w_2}})...>
\nonumber \\
\lbrace{C_{{\lbrack}k-3|l-3\rbrack}}
\alpha_{{\lbrack}-3|k-3\rbrack}\alpha_{{\lbrack}-3|n+6-k-l\rbrack}
\alpha_{{\lbrack}-3|n+6\rbrack}
+
{C_{{\lbrack}k-3|n+6-k-l\rbrack}} 
\alpha_{{\lbrack}-3|k-3\rbrack}\alpha_{{\lbrack}-3|l-3\rbrack}
\alpha_{{\lbrack}-3|n+6\rbrack} ~~~~~~~~~~~   
\nonumber \\
+
{C_{{\lbrack}k-3|n+6-k-l\rbrack}}
\alpha_{{\lbrack}-3|l-3\rbrack}\alpha_{{\lbrack}-3|n+6-k-l\rbrack}
\alpha_{{\lbrack}-3|n+6\rbrack}\rbrace ~~~~~~~~~~~~~~
\eea
Again, it is easy to see that in the absence of the ghost-matter
mixing
($\alpha=1$, all C are  picture-independent)
this contribution would sum up to
\be 
{1\over2}C\lambda^3log\Lambda
\int{d^2w_1}\int{d^2w_2} V(w_1,\bar{w_1}) 
V(w_2,\bar{w_2}) 
\ee  
precisely cancelling the divergence of the same $\lambda^3$-type,
originating from the renormalization of the $\lambda\varphi$ cross term
under the flow. In the picture-independent case this insures
that the renormalization (11) of the dilaton field under the flow
does  not bring about any additional singularities from higher order
terms, such as the
cubic one and the $\lambda\varphi$ cross-term.
In particular, this guarantees that terms
of the type 
\be
{\sim}C\lambda^3{log\Lambda}\int_\Lambda{d^2w}V_5(w,\bar{w})
\ee
( $\int_\Lambda$ denotes the world sheet integral cut at the 
$\Lambda$ scale)
never appear in the dilaton or other perturbative close string field
beta-functions in the picture-independent case.
This insures, in turn, that in the absence of the ghost-matter
mixing the world sheet beta-function is always deterministic
(just as it is well-known to be the case in 
the standard string perturbation theory)
On the contrary, should the terms of this 
type appear in the beta-function, that would imply that
the RG equations become stochastic, since from the
point of view of the space-time fields,
world sheet operator  $\int_\Lambda{d^2w}V_5(w,\bar{w})$
is a stochastic random variable, with the cutoff parameter $\Lambda$
playing the role of the stochastic time.
In this case, the RG equations have the form of non-Markovian Langevin
equations where the memory of the noise is determined by the
world sheet correlation of the $V_5$-operators.
This exactly is what happens in the ghost-matter mixing backgrounds,
when the OPE of vertex operators are picture-dependent.
As a result of the picture-dependence, the flows of
the $\lambda\varphi$ and $\lambda^3$ terms under the RG do not cancel
each other and, as a result, the beta-function of the dilaton
gets the stochastic terms, as $A_{\lambda^3}+A_{\lambda\varphi}\neq{0}$
Indeed, using the identities (7) we can write (15) in the form:

\bea
{1\over2}{C_{{\lbrack}-3|-3\rbrack}}
\alpha_{{\lbrack}-3|1\rbrack}
\lambda^3{log}\Lambda{\sum_{n=0}^{\infty}}
3^{-n-9}\sum_{k,l=0;k{\neq}2,3;l{\neq}2,3;k+l{\neq}n+5,n+6}^{k+l=n+9}
{{(n+9)!}\over{k!l!(n+9-k-l)!}}
\nonumber \\
\int{d^2w_1}\int{d^2w_2}<V_\varphi^{(-6)}(w_1, 
\bar{w_1})V_5^{(n+6)}(w_2,\bar{w_2})...>
\eea
giving the contribution of the $A_{\lambda^3}$-term to the flow.
Now, to get the total flow on the $\lambda^3$ level
one has to subtract the flow
 (14) from this expression.
The difference would give the overall coefficient before
the stochastic term of the type (17) in the beta-function of the dilaton.
Comparing (14) and (18) and performing the summations
 we find that the additional renormalization of the 
dilaton field, necessary to remove the  extra singularities from
the $\lambda^3$ and $\lambda\varphi$ terms, arising due to the
OPE picture dependence in ghost-matter mixing backgrounds,
is given by
\bea
\varphi\rightarrow\varphi-
{C_{{\lbrack}-3|-3\rbrack}} 
\alpha_{{\lbrack}-3|1\rbrack} \lambda^3{log}\Lambda
\int{d^2w}V_5^{(-3)}(w,\bar{w})
\nonumber \\
\times\lbrace -1 + \sum_{n=0}^{\infty}\lbrack
(n+4)^2(n+5)^3(n+6)2^{-2n-12}
+{1\over{96}}(n+8)(n+9)(n+13)({2\over3})^{n+9} -
\nonumber \\
((n+4)^2(n+5)+(1/2)(n+5)^2(n+6)){2^{-n-6}}\rbrack
\rbrace
\eea
The summation over n converges to
\bea
-\sigma= -1 +
\sum_{n=0}^{\infty} \lbrack
(n+4)^2(n+5)^3(n+6)2^{-2n-12} +
{1\over{96}}(n+8)(n+9)(n+13)({2\over3})^{n+9} 
\nonumber \\
- 
((n+4)^2(n+5)+(1/2)(n+5)^2(n+6)){2^{-n-6}}\rbrack
=-1,534.
\eea
Therefore the resulting beta-function equations
for the dilaton in the ghost-matter mixing background give:
\be
{{d\varphi}\over{d{log}\Lambda}}=-{{\delta{S_\varphi}}\over{\delta\varphi}}
+\sigma{C_{{\lbrack}-3|-3\rbrack}}
\alpha_{{\lbrack}-3|1\rbrack}\int_\Lambda{d^2w}V_5^{(-3)}(w,\bar{w})
\ee
where $S_\varphi\sim\int{dx}\partial_m\varphi\partial^m\varphi$
is the low energy effective action for the dilaton in the absence
of the $V_5$ background.

With the restored momentum dependence, this
equation can be written as 

\bea 
{{d\varphi(p)}\over{d{log}\Lambda}}=
-{{\delta{S_\varphi}}\over{\delta\varphi(p)}}
+\sigma C(p) \int d^4 q \lambda(q)\int_\Lambda{d^2w}V_5^{(-3)}(w, q) 
\nonumber
\\
 C(p) = \int d^4 k  {C_{{\lbrack}-3|-3\rbrack}}(p,k)  
\lambda({{k+p}\over2})
\lambda({{k-p}\over2})
 \label{Langevin}
\eea 
with $\sigma=1,534...$
So far we were considering only one particular example of the
ghost-matter mixing - the brane-like vertex operator
$V_5^{(-6)}$ and its incarnations in higher pictures.
There are other examples of vertex operators with
 ghost-matter mixing, and they also lead to
stochastic terms in the beta-function of the dilaton.
In particular, we have also considered the dilaton field
in the background of closed string operators of higher ghost cohomologies:
\be
W_5\sim\int{d^2z} e^{-4\phi-\bar\phi}\partial X_{(m_1}...\partial
X_{m_3 )}
\bar\psi_{m_6}e^{ik^\perp{X}}G_{m_1...m_5m_6}
\ee
where the G-tensor is symmetric and traceless in $m_1$,...$m_5$
(round brackets imply the symmetrization in space-time indices)
and $k^\perp$ is  transversely to $m_1,...m_6$ directions
and
\be
U_5\sim\int{d^2z}{e^{-4\phi-\bar\phi}}\partial X_{(m_1 }...\partial 
X_{m_4)}
\psi_{m_5}\psi_{m_6}\bar\psi_{m_7}e^{ik^\perp{X}}G_{m_1...m_7}
\ee
We have found that, even  though the OPE details are quite different
in each case, nevertheless in the end one always gets
the beta-function equations in the form (22).
The crucial point is that the $\sigma$ factor,
reflecting the stochasticity of the beta-function,
  appears to be universal and its value is independent on details
of the ghost-matter mixing.
Namely, we have found \cite{progress} 
$\sigma=1,541...$ for the $W_5$ insertion and $\sigma=1,538...$
for the of $U_5$ case .
In other words, the form of these equations is determined by
 the corresponding coefficient of the OPE of two operators before
the dilaton, 
and the numerical factor of
$\sigma$  
What is quite remarkable, the coefficient 
$\sigma=1,534...$ before the stochastic term is always the same
and seems to be invariant on details of the ghost-matter mixing
(in the absence of the ghost-matter mixing, of course $\sigma=0$).
What is even more remarkable, can easily check that in fact 
\be
\sigma=\ln \delta 
\ee
where $\delta=4,669...$ is the famous Feigenbaum universality
constant describing the universal scaling of the 
iteration parameter in a huge variety of dynamical systems
under bifurcations and
transitions from order to chaos \cite{Feigenbaum:ys}.
Usually this constant can be obtained numerically for
the dynamical systems under the bifurcations.
In our approach, however, the log of the 
Feigenbaum constant has emerged analytically, as the limit of convergent
series (20)  in the stochastic term in the beta-function for
various ghost-matter mixing backgrounds.
In the next section we will show that the appearance of the Feigenbaum
constant in the beta-function for the dilaton
in ghost-matter mixing backgrounds
is not at all occasional, 
but is
deeply related to  the peculiarities of the non-Markovian stochastic
processes, associated with the Langevin equations and their
implications for the space-time geometry.

\section{Feigenbaum universality and fixed points of stochastic RG 
equations}

To understand the physical meaning behind
the appearance of the Feigenbaum constant in (20), it is necessary to 
analyze
the non-Markovian Fokker-Planck (FP) equation describing the stochastic
process which can be straightforwardly derived from the Langevin
equation  (\ref{Langevin}).
We  shall present here FP equation for scaling functions $\lambda(q) =
\lambda_0/q^4$
\bea
{{\partial{P_{FP}}(\varphi,\tau)}\over{\partial\tau}}
=-
\int{{d^4p}}\int{{d^4q}}
{{\delta}\over{\delta\varphi(p,\tau)}}
({{\delta{S_\varphi}}\over{\delta\varphi(q,\tau)}}
P_{FP}(\varphi,\tau))
\nonumber \\
+\sigma^2\lambda_0^6
\int{d^4k_1}\int{d^4k_2}\int{{d^4p}\over{p^4}}\int{{d^4q}\over{q^4}}
\int{d\xi}
\nonumber \\
\alpha_{{\lbrack}-3|1{\rbrack}}C_{{\lbrack}-3|-3{\rbrack}}
({{k_1+p}\over2}) 
\alpha_{{\lbrack}-3|1{\rbrack}}C_{{\lbrack}-3|-3{\rbrack}}({{k_2+q}\over2}) 
{{\delta}\over{\delta\varphi(p,\tau)}}G_5(\xi,\tau)
{{\delta}\over{\delta\varphi(q,\xi)}}
P_{FP}(\varphi,\tau)
\eea
where 
$\tau=log\Lambda$
 now plays the role of the 
 stochastic time variable. The Green's function $G_5(\xi,\tau,p,q)$
is defined by the cutoff dependence of the two-point correlator
of the $V_5$-vertices:
\bea
G_5(\xi,\tau)=\int_{\Lambda_1}d^2z\int_{\Lambda_2}{d^2w}
|z-w|^{-4}\delta(p+q)=
({{1+e^{\xi-\tau}}\over{1-e^{\xi-\tau}}})^2\delta(p+q), 
\xi=log\Lambda_1,\tau=log\Lambda_2
\eea
We shall look  for the anzats solving this equation in the form (for
more details see \cite{Polyakov:2000xc} and references therein):
\bea
 {P_{FP}(\varphi,\tau)}=\exp[-{H_{ADM}}(\varphi,\tau)] =
\exp[-\int{d^4p}\lbrace{g(\tau)(\partial_\tau\varphi)^2
+f(\tau)p^2\varphi^2}\rbrace ]
\eea
Substituting it into (26) we find that (28) solves  the Fokker-Planck 
equation
provided that the functions $f(\tau)$ and $g(\tau)$ satisfy the 
following differential
equations:
\bea
g^\prime(\tau)+4g(\tau)+\frac{\sigma^2\lambda_0^6}{2}=0
\nonumber \\
{1\over4}f^{\prime\prime}+(1+{1\over{4\tau}})f^{\prime}
+(1+{1\over{4\tau}}+{1\over{4\sigma^2\lambda_0^6}}(1-{1\over{\tau^2}}))f
-(1-{1\over{\tau^2}})(e^{-2\tau}+{1\over{4\sigma^2\lambda_0^6}})=1
\eea
The first equation is elementary, its solution is given by
\bea
g(\tau)=\frac{\sigma^2\lambda_0^6}{2}(e^{-4\tau}-1), ~~~ \tau <0
\eea
The second equation on $f(\tau)$ can be reduced to the Bessel type
equation by substituting $$f(\tau)=\rho(\tau)e^{-2\tau}+
{1\over{\sigma^2\lambda_0^6}}$$
The solution is given by
\be
f(\tau)=1+\sigma^2\lambda_0^6e^{-2\tau}(1+{J_{1\over{\sigma\lambda_0^3}}}
({{\tau}\over{\sigma\lambda_0^3}}))
\ee
where 
$
{J_{1\over{\sigma\lambda_0^3}}}
({{\tau}\over{\sigma\lambda_0^3}})$ is the Bessel's function.
In terms of the $\tau$ coordinate,  the stochastic process,
describing the RG flow in ghost-matter mixing backgrounds,
evolves  in the direction of $\tau=-\infty$.
Next, let us study the behaviour of the Fokker-Planck
distribution (28), (30), (31)
in the conformal limit of $\tau\rightarrow{-\infty}$
In this limit the exponents become very large and moreover
\be
{J_{1\over{\sigma\lambda_0^3}}}
({{\tau}\over{\sigma\lambda_0^3}})\sim{O}({1\over{\sqrt{\tau}}})<<1
\ee
and after rescaling the distribution reduces to
\be
H(\varphi,\tau)=R^2\int{d^4p}{\lbrace}e^{-4\tau}(\partial_\tau\varphi)^2
+p^2e^{-2\tau}\varphi^2\rbrace
\ee
which is just the ADM Hamiltonian for the $AdS_5$ gravity
in the temporal gauge \cite{verl}. 
It is easy to see that the $\lambda_0^6$ parameter
has the meaning of the square of the radius $R^2$
of the metric.

Let us now analyze in more details
 the solution (28), (30), (31) of the non-Markovian FP equation,
leading to the new space geometry.
 Let us note first of all that the limit $\lambda_0 \rightarrow 0$ is 
 not the same as  $\lambda_0=0$ (ghost-matter mixing absent).
The RG flow described by the effective metric
(30),(31) must be single-valued; since  Bessel's functions at zero 
argument
are single-valued for the integer orders only,
this leads to the quantization condition:
\be
(\sigma\lambda_0^3)^{-1}=N
\ee
Moreover, since $J_\nu(\tau)\sim{\tau^\nu}$
 as $\tau\rightarrow{0}$, the absence of unphysical singularities at 
$\tau=0$ requires N to be positive. 
The quantization condition (32)
implies that
\be
((\lambda_0)_N)^{-3}=N\sigma, ~~~~~~~
e^{(\lambda_0)_N^{-3}}=\delta^{N}
\ee
implying the iteration law:
\be
{{{e^{(\lambda_0)_{N+1}^{-3}}}-e^{(\lambda_0)_{N}^{-3}}}
\over{{e^{(\lambda_0)_{N}^{-3}}}-e^{(\lambda_0)_{N-1}^{-3}}}}
=\delta
\ee
with $\delta$ being the Feigenbaum number.

Therefore the Feigenbaum iteration rule (36)   determines the scaling
of characteristic curvatures of geometries emerging at
the fixed points of the stochastic renormalization group.
The role of iteration parameter characterizing the bifurcations
is played by $\sim{e^{-{1\over{R^2}}}}$, vanishing at $R=0$ and being 
finite
at large R, as it should be the case for the scaling parameter of 
the Feigenbaum iteration scheme.

From the quantization condition (34) it is clear that the stochastic
renormalization group
(22) has  fixed points exists
for $0<\lambda_0<1$, (physically, these points correspond
 to {\it large} curvatures). Moreover  the period doublings 
that lead to the transition to chaos corresponds to $N \rightarrow
\infty$, i.e.
 $\lambda_0 \rightarrow 0$, which is a singularity. So we reached an
amazing conclusion that precisely near singularity our RG flow becomes 
chaotic. It is tempting to assume that this may be the mechanism which 
can solve the problem of singularities in string theory.

\section{Conclusion}

 In this Letter we discussed how  matter-ghost mixing can radically modify
the nature
 of the world sheet RG flows and lead to the emergence of chaos near 
 curvature singularities.   Here 
we  analyzed  only  dilaton evolution, but the similar picture  
 can be  obtained for other massless fields, for example metric
 \cite{progress}.

  It is    amusing that
recently chaotic behaviour of metric  was discussed in 
\cite{damour} (for earlier papers see \cite{Ivashchuk:1999rm} and
references therein)
where  the emergence of chaos in supergravity near cosmological 
singularity was
 demonstrated in the presence 
 of  higher rank antisymmetric tensor fields, i.e. R-R fields.
  It will be extremely interesting to
understand   how 
 chaos emerging during cosmological evolution in supergravity can be  
 related to the chaotic nature of RG flows in underlying string theory in
the presence
  of the sources of the background R-R fields. 

It is tempting to assume that the resolution of the singularities
problem is   transition to chaos  and emergence of smooth
distributions of fields, not restricted on-shell.
One can  imagine that curvature $R$ is some  new  
"Reynolds" number in string theory 
 and for large $R$    one 
have transition 
to chaotic behaviour in a similar fashion like in hydrodynamics 
there is a transition from a laminar to a turbulent flow. These ideas
 definitely need further investigation.

I.K. is supported in part by PPARC
rolling grant PPA/G/O/1998/00567 and  EC TMR grants
HPRN-CT-2000-00152 and  HRRN-CT-2000-00148.
D.P. acknowledges the support of the Academy of Finland 
under the Project no. 54023 and both I.K. and D.P. acknowledge
interesting discussions with T. Damour and  the
 hospitality of 
Institute des Hautes Etudes Scientifiques (IHES) in
Bures-sur-Yvette where part of this work has been done.


\begin{thebibliography}{99}

\bibitem{books}
 A.M.Polyakov, "Gauge Fields and Strings", Harwood Academic Publishers, 
(1987).\\
 M.B.Green, J.H.Schwarz and E.Witten, "Superstring Theory", vol 1,2 (CUP), 
(1987).\\
J. Polchinski, ``String Theory'', vol 1,2 (CUP), (1998);

\bibitem{rg} 
D.Fiedan , Phys.Rev.Lett. {\bf 45}, (1980), 1057; Ann.of Phys.(N.Y.),
{\bf 163}, (1985), 318. \\
E.S.Fradkin and A.A.Tseytlin, Nucl.Phys. {\bf B 261}, (1985), 1. \\
C.G.Callan, D.Friedan, E.J.Martinec and M.J.Perry, Nucl.Phys. {\bf B
262}, (1985), 593. \\
C.G.Callan and Z.Gan, Nucl.Phys. {\bf B 272}, (1986), 647.\\
A.B.Zamolodchikov, JETP.Lett.{\bf 43} (1986),730; 
Sov.J.Nucl.Phys.{\bf 46(6)} (1987),
1090. 




\bibitem{Polyakov:2000xc}
D.~Polyakov,
Class.\ Quant.\ Grav.\  {\bf 18} (2001) 1979
[arXiv:hep-th/0005094].

\bibitem{Kogan:2000nw}
I.~I.~Kogan and D.~Polyakov,
Int.\ J.\ Mod.\ Phys.\ A {\bf 16}, 2559 (2001)
[arXiv:hep-th/0012128].


\bibitem{Polyakov:2001zr}
D.~Polyakov,
Phys.\ Rev.\ D {\bf 65}, 084041 (2002)
[arXiv:hep-th/0111227].

\bibitem{Kogan:2002tn}
I.~I.~Kogan and D.~Polyakov,
Int.\ J.\ Mod.\ Phys.\ A {\bf 18} (2003) 1827
[arXiv:hep-th/0208036].


\bibitem{Feigenbaum:ys}
M.~J.~Feigenbaum,
J.\ Statist.\ Phys.\  {\bf 19} (1978) 25.



\bibitem{progress}
I.~I.~Kogan and D.~Polyakov,
to be published

\bibitem{verl}
J.~de Boer, E.~Verlinde and H.~Verlinde,
JHEP {\bf 0008} (2000) 003
[arXiv:hep-th/9912012].



\bibitem{damour}
T.~Damour, M.~Henneaux,
Phys.\ Rev.\ Lett.\  {\bf 85} 920, (2000)
hep-th/0003139;Phys.\ Rev.\ Lett.\  {\bf 86}  4749 (2001)  hep-th/0012172;
Gen.\ Rel.\ Grav.\  {\bf 32} (2000) 2339. \\ 
T.~Damour, M.~Henneaux, B.~Julia and H.~Nicolai,
Phys.\ Lett.\ B {\bf 509} (2001) 323
[arXiv:hep-th/0103094].\\
T.~Damour, M.~Henneaux and H.~Nicolai,
Phys.\ Rev.\ Lett.\  {\bf 89} (2002) 221601
[arXiv:hep-th/0207267]. \\
T.~Damour,
Int.\ J.\ Mod.\ Phys.\ A {\bf 17} (2002) 2655.


\bibitem{Ivashchuk:1999rm}
V.~D.~Ivashchuk and V.~N.~Melnikov,
J.\ Math.\ Phys.\  {\bf 41} (2000) 6341
[arXiv:hep-th/9904077].






\end{thebibliography}
\end{document}